\documentclass[runningheads]{llncs}

\usepackage[T1]{fontenc}
\usepackage{amsmath,amssymb}
\usepackage{graphicx}
\graphicspath{{figures/}}
\usepackage{booktabs}
\usepackage{hyperref}
\usepackage{url}

\IfFileExists{microtype.sty}{\usepackage[protrusion=true,expansion=false]{microtype}}{}

\begin{document}

\title{Civilizational Metamaterials: Engineering Coordination Under Capability Gradients and Structural Turbulence}
\titlerunning{Civilizational Metamaterials}

\author{David Orban\orcidID{0009-0004-4954-1147}}

\authorrunning{D. Orban}

\institute{Independent Researcher\\
\email{david@davidorban.com}}

\maketitle

\begin{abstract}
We argue that governance must transition from a normative discipline to an engineering discipline, and we develop a formal framework, inspired by the physics of metamaterials, to make this transition quantitative and testable. Artificial General Intelligence affects civilization primarily by increasing decision velocity while human verification capacity remains bounded. When the cost of validating AI-generated outputs exceeds the expected utility of acting on them, rational agents default to inaction: a stable but catastrophic Nash equilibrium we term the \emph{Freezing Equilibrium}. Drawing on metamaterials, where emergent macro-properties arise from designed microstructure, we develop a phenomenological constitutive law for institutional coordination: $R_{\mathrm{eff}} = \beta \cdot (1-\rho) \cdot (1-\tau) \cdot (1-\gamma \rho \tau)$, where $\beta$ is the decision branching factor, $\rho$ is provenance fidelity, $\tau$ is the verification rate, and $\gamma \in [0,1]$ captures correlated-detection synergy between provenance and verification failures. The model predicts a sharp phase transition between self-healing ($R_{\mathrm{eff}} < 1$) and self-destabilizing ($R_{\mathrm{eff}} > 1$) regimes. We introduce a three-class provenance taxonomy: cryptographic, institutional, and \emph{context binding}, and derive four falsifiable hypotheses with a proposed 12-week stepped-wedge cluster-randomized trial in government grant review panels. The framework bridges AI alignment theory and institutional design.

\keywords{AGI governance \and coordination theory \and institutional design \and decision provenance \and AI safety \and multi-agent systems}
\end{abstract}

\section{Introduction}
\label{sec:intro}

Previous waves of information technology accelerated transmission: the telegraph collapsed geographic latency; the internet democratized access. Artificial General Intelligence accelerates something qualitatively different: \emph{synthesis and decision}. The specific risk is not speed alone but the structural consequence of decision velocity outpacing verification velocity. Without engineered countermeasures, the cost of verifying reality exceeds the cost of generating fiction, and institutions enter a regime where rational inaction becomes the dominant strategy.

\begin{figure}[t]
\centering
\includegraphics[width=\textwidth]{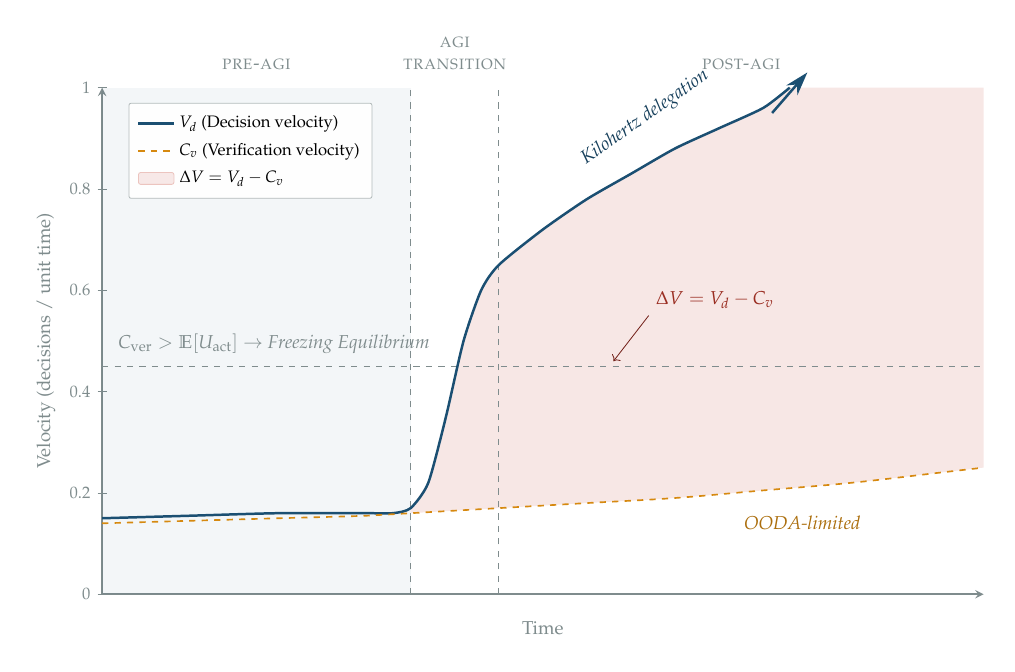}
\caption{The Decision--Verification Gap. Decision velocity $V_d$ diverges from verification velocity $C_v$ as AGI accelerates delegation beyond human verification capacity. The shaded region represents unverified decisions accumulating faster than they can be processed. When $C_{\mathrm{ver}} > \mathbb{E}[U_{\mathrm{act}}]$, the system reaches the Freezing Equilibrium.}
\label{fig:gap}
\end{figure}

\paragraph{The decision--verification gap.}
Let $V_d$ denote the rate at which AI systems generate decisions and $C_v$ the rate at which human (or human-supervised) processes can verify them. Historically, both were bounded by cognitive throughput, and $V_d \approx C_v$. AGI decouples these rates: synthetic principals execute directives at kilohertz frequencies while human verification remains tethered to the biological orientation phase of the OODA loop~\cite{boyd1996essence}, requiring 0.2--2.0\,s per assessment~\cite{card1983psychology}. Empirical analysis of AI capability benchmarks suggests that $V_d$ is growing not exponentially but \emph{super}exponentially\,---\,with a positive third derivative, a ``jolt''\,---\,implying that the velocity gradient $\Delta V = V_d - C_v$ itself accelerates over time rather than merely growing~\cite{orban2025jolting}. Unverified claims therefore accumulate faster than they can be processed (Fig.~\ref{fig:gap}).

\paragraph{The Freezing Equilibrium.}
When the expected cost of verification exceeds the expected utility of action,
\begin{equation}
C_{\mathrm{ver}} > \mathbb{E}[U_{\mathrm{act}}] \implies \text{Stasis},
\label{eq:freeze}
\end{equation}
rational agents wait. We formalize as follows. Each agent chooses among three actions\,---\,ACT-blind, ACT-verified, or WAIT\,---\,in response to a claim that is legitimate with prior probability $p$, where $p$ falls as $\Delta V$ grows. Payoffs are
\begin{align*}
u(\text{ACT-blind}) &= p\,U_{\mathrm{act}} - (1-p)\,L, \\
u(\text{ACT-verified}) &= p\,U_{\mathrm{act}} - C_{\mathrm{ver}}, \\
u(\text{WAIT}) &= 0,
\end{align*}
where $L$ is the loss from acting on a false claim. The all-WAIT profile is a strict Nash equilibrium when verification is unaffordable, $C_{\mathrm{ver}} > p\,U_{\mathrm{act}}$, and blind action is too risky, $(1-p)\,L > p\,U_{\mathrm{act}}$. Both inequalities tighten with $\Delta V$. The equilibrium is socially catastrophic and dominated by profiles in which agents share verification cost via institutional infrastructure\,---\,exactly what $\rho$ and $\tau$ in the constitutive law (Section~\ref{sec:constitutive}) operationalize.

\emph{Worked example.} An environmental regulatory agency historically processed 200 permit applications per year at $\sim$40 analyst-hours of verification each. An AGI-augmented consulting industry now submits 2{,}000 assessments annually of comparable apparent quality, with no proportional scaling of verification capacity. Analysts, unable to distinguish rigorous from confabulated assessments without full review, face inequality~\eqref{eq:freeze}: the expected verification cost exceeds the expected utility of approval, since approving an unverified fraudulent assessment carries career and legal risk. Decisions defer. Development stalls\,---\,not from regulatory opposition, but from verification paralysis. The Freezing Equilibrium is the predictable outcome of a verification bottleneck under high $\Delta V$, not a failure of will.

\paragraph{The metamaterial analogy as theory.}
Engineered metamaterials\,---\,periodic microstructures that produce emergent macro-properties such as photonic bandgaps, mechanical anisotropy, and phase transitions\,---\,offer more than a metaphor. The analogy generates specific, falsifiable predictions that a generic ``governance framework'' would not.

We distinguish the four hypotheses by analogical-import strength. H1 (bandgap) and H2 (anisotropy) are \emph{structural imports}: the periodicity-forbids-propagation argument (H1) and the directional decomposition of $R_{\mathrm{eff}}$ (H2) draw substance from metamaterial-physics formalism. H3 (threshold-crossing superadditivity) follows from the correlated-detection term and is properly a branching-process prediction. H4 (hysteresis) is a \emph{heuristic import}: hysteresis appears widely in dynamical systems and the metamaterial framing motivates the asymmetry without deriving it. Empirical failure of H1 or H2 falsifies the structural framing; failure of H3 or H4 falsifies the constitutive law or its analogical packaging respectively. Only H1 and H2 carry the burden of justifying metamaterial language.

\paragraph{Contribution.}
The treatment of governance rules as designable microstructure extends Ostrom's institutional design tradition~\cite{ostrom1990governing} by providing a formal threshold criterion ($R_{\mathrm{eff}} < 1$) for when the institutional design is sufficient.

This paper makes four contributions: (1)~a phenomenological constitutive law for institutional coordination, parameterized by designable features, with a sharp phase transition (Section~\ref{sec:constitutive}); (2)~a three-class provenance taxonomy identifying \emph{context binding} as the missing third class (Section~\ref{sec:provenance}); (3)~treatment of AI agents as ``synthetic principals'' requiring distinct governance primitives (Section~\ref{sec:synthetic}); (4)~four falsifiable hypotheses with a concrete experimental design (Section~\ref{sec:empirical}).

\section{The Constitutive Law and Phase Transition}
\label{sec:constitutive}

The model belongs to the family of stochastic branching processes~\cite{anderson1991infectious,hethcote2000mathematics} and cascading failure analysis~\cite{watts2002simple,buldyrev2010catastrophic,dobson2024cascading}. We contribute an \emph{institutional parameterization}: (i)~$\beta$ is a \emph{design variable} under institutional control, not an exogenous rate; (ii)~the effective reproduction rate decomposes into actionable targets ($\rho$, $\tau$) with a synergy term ($\gamma$); and (iii)~the provenance taxonomy (Section~\ref{sec:provenance}) makes these parameters measurable.

\subsection{Failure Propagation as a Branching Process}

Institutional stability is a structural property dictated by the rate at which errors propagate across decision nodes. Each node in a decision network acts as either a signal attenuator or a failure amplifier. We model this as a stochastic branching process parameterized by four quantities:

\begin{itemize}
\item \textbf{Branching factor} $\beta$: the average number of downstream nodes a single decision impacts. This is \emph{endogenous}: rate limits, delegation boundaries, and dual-control requirements reduce $\beta$ directly.
\item \textbf{Provenance fidelity} $\rho \in [0,1]$: the probability that the source and transformation history of information is cryptographically bound to the decision unit.
\item \textbf{Verification rate} $\tau \in [0,1]$: the probability that a node detects and halts an erroneous claim.
\item \textbf{Correlated-detection coefficient} $\gamma \in [0,1]$: an interaction term capturing that provenance and verification target overlapping failure modes. When both controls are active, errors that would defeat one check are correlated with errors that would defeat the other; the joint failure probability is therefore \emph{lower} than the independent baseline $(1-\rho)(1-\tau)$ by an amount controlled by $\gamma$. $\gamma = 0$ recovers the independent multiplicative model; $\gamma = 1$ is the maximum-correlation limit.
\end{itemize}

We propose the following phenomenological constitutive law for the effective failure propagation rate:
\begin{equation}
R_{\mathrm{eff}} = \beta \cdot (1-\rho) \cdot (1-\tau) \cdot (1-\gamma \rho \tau).
\label{eq:reff}
\end{equation}

$R_{\mathrm{eff}}$ is a \emph{phenomenological ansatz}: its justification is empirical, like Hooke's law or Ohm's law. The multiplicative structure $(1-\rho)(1-\tau)$ reflects sequential filtering under the assumption of independent failures: at each node, an error must survive both a provenance check and a verification check, and the two checks are assumed independent. In practice they are not: an actor who can defeat the provenance system is more likely to be able to defeat verification (and vice versa) because both systems are partly targeting the same underlying capability gap. The correction term $-\gamma\rho\tau$ models this correlation, and $\gamma \in [0,1]$ ensures $R_{\mathrm{eff}} \geq 0$ for all admissible parameter values.

The bilinear form $\gamma\rho\tau$ is the lowest-order non-trivial interaction that vanishes when either $\rho$ or $\tau$ is zero. The qualitative threshold-crossing prediction (H3, Section~\ref{sec:empirical}) is robust to the functional form; quantitative critical thresholds vary with the specific interaction and must be empirically calibrated (Section~\ref{sec:sensitivity}). A first-principles derivation from a node-level correlation model is left for future work.

\subsection{The Phase Transition}

The system exhibits a sharp phase transition at $R_{\mathrm{eff}} = 1$. This threshold property is inherited from the branching-process model class~\cite{anderson1991infectious,hethcote2000mathematics} and is not, in itself, a novel prediction:

\begin{itemize}
\item \textbf{Damped regime} ($R_{\mathrm{eff}} < 1$): Errors decay exponentially with network depth. The system is self-healing; tail risk is bounded.
\item \textbf{Turbulent regime} ($R_{\mathrm{eff}} > 1$): Errors amplify exponentially. The system is self-destabilizing; cascade depth follows a power-law distribution with fat tails.
\end{itemize}

What the institutional parameterization adds is that the sub-critical condition $R_{\mathrm{eff}} < 1$ can be \emph{engineered} rather than merely observed: $\beta$ can be constrained by delegation policy, $\rho$ can be increased by provenance infrastructure, and $\tau$ can be raised by verification protocols. The constitutive law makes the relationship between these design choices and system-level stability quantitative.

The critical verification threshold for a given branching factor is $\tau^* = 1 - 1/\beta$ (in the simplified case where $\rho = 0$ and $\gamma = 0$). For a standard hierarchical panel with $\beta = 10$, stability requires $\tau > 0.90$. As AGI-accelerated delegation pushes the effective branching factor higher\,---\,in projected agentic-AI regimes where a single human authorization can trigger order-of-magnitude more downstream actions\,---\,the required verification fidelity approaches limits that legacy institutions cannot sustain without automated scaffolding. The specific value of $\beta$ under agentic delegation is an empirical question; for illustration, $\beta = 50$ would require $\tau \approx 0.98$.

\begin{figure}[t]
\centering
\includegraphics[width=\textwidth]{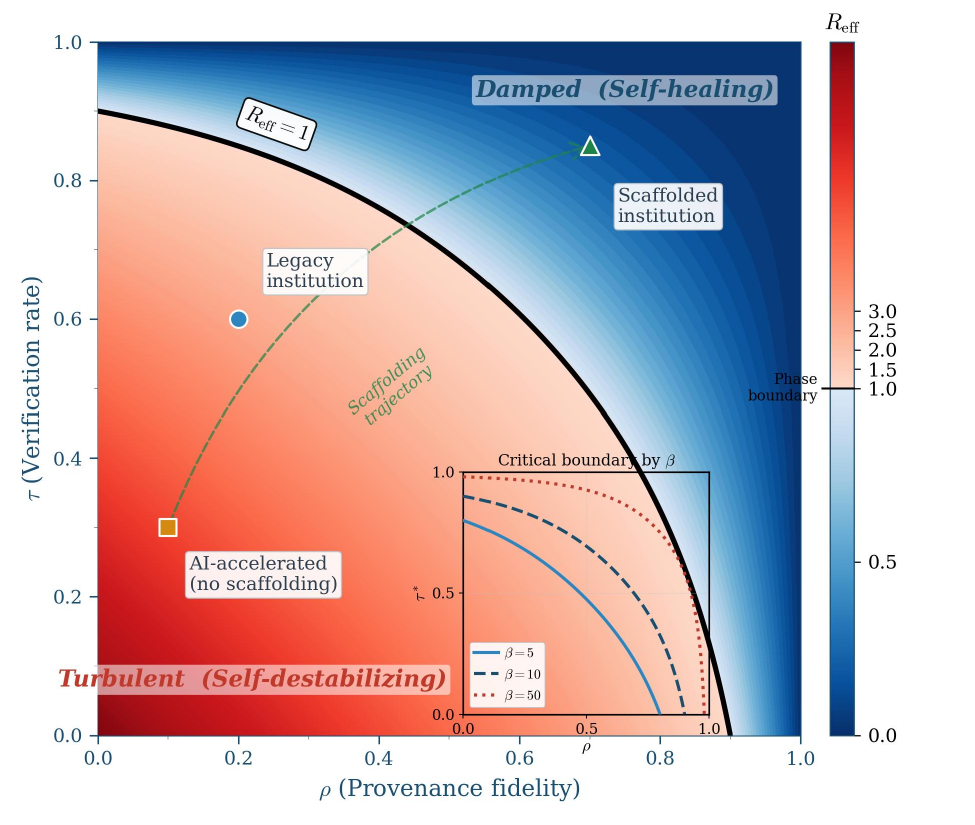}
\caption{Phase transition diagram for $R_{\mathrm{eff}}$ in the $(\rho, \tau)$ parameter space ($\beta=10$, $\gamma=1$). The bold contour marks the critical boundary $R_{\mathrm{eff}}=1$; blue region is damped (self-healing), red is turbulent (self-destabilizing). Three archetypal institutional positions are shown. Inset: boundary shift under varying branching factor $\beta$.}
\label{fig:phase}
\end{figure}

The correlated-detection term $(1 - \gamma \rho \tau)$ produces the framework's key threshold-crossing prediction. Because $\gamma \rho \tau$ grows multiplicatively with both controls, the combined high-$\rho$, high-$\tau$ regime can satisfy $R_{\mathrm{eff}} < 1$ at parameter values where either intervention alone leaves the system turbulent. A small coordinated improvement in both $\rho$ and $\tau$ can flip the system from turbulent to damped (Fig.~\ref{fig:phase}); equivalent improvements in either alone, while reducing $R_{\mathrm{eff}}$, do not in general cross the critical boundary. This threshold-crossing behavior\,---\,rather than the magnitude of joint reduction relative to the sum of singletons\,---\,is the operational meaning of ``superadditive interventions'' in what follows.

Crucially, $\beta$ in institutional networks is itself a \emph{design parameter}, not the exogenous rate of an epidemiological process. Rate limits bound the maximum $\beta$; delegation boundaries partition the network into sub-graphs; dual-control requirements reduce the effective $\beta$ for high-impact decisions. Scaffolding therefore attacks all three terms in~\eqref{eq:reff} simultaneously: reducing $\beta$, increasing $\rho$, and increasing $\tau$. The coordination ``bandgap''\,---\,the range of failure modes that cannot propagate\,---\,depends on the combination of these design choices, in analogy to how a metamaterial's bandgap width depends on lattice periodicity and contrast ratio.

\section{A Three-Class Provenance Taxonomy}
\label{sec:provenance}

Current scaffolding initiatives focus on two provenance mechanisms: content provenance (C2PA~\cite{c2pa2024}) and identity verification (Proof of Personhood). Neither addresses the full attack surface. We propose three complementary classes.

\subsection{Class A: Cryptographic Provenance}

Cryptographic provenance establishes the chain of custody from origin to current state via signatures that are computationally infeasible to forge. The C2PA Technical Specification~2.0 embeds tamper-evident manifests directly within asset bitstreams using JUMBF structures, including hash assertions for content binding and action assertions for transformation history~\cite{c2pa2024}.

\emph{Failure mode}: Key compromise, implementation bugs, or capture of the signing infrastructure. \emph{Mitigation}: Hardware security modules, key rotation, distributed signing, and audit of signing infrastructure.

\subsection{Class B: Institutional Provenance}

A document signed by a ministry has institutional provenance: the reputation of the signing entity provides assurance beyond the cryptographic fact. The IETF SCITT standard~\cite{birkholz2025scitt} operationalizes this by submitting signed statements to a Transparency Service that issues receipts anchored in an append-only Merkle tree, preventing equivocation: an institution cannot repudiate a decision without invalidating the cryptographic proof held by downstream parties.

\emph{Failure mode}: Institutional corruption, capture, or incompetence. The signature is valid but the institution's judgment is compromised. \emph{Mitigation}: Separation of powers, external audit, competitive verification markets, and reputation tracking.

A critical insight: high cryptographic provenance with low institutional provenance (a corrupt institution with perfect signatures) is more dangerous than the reverse, because the cryptographic validity provides false assurance.

\subsection{Class C: Context Binding (Novel Contribution)}
\label{sec:contextbinding}

Classes A and B secure integrity and attribution but fail against \emph{``Valid Credential, Invalid Context''} attacks, where adversaries replay authorized outputs outside their intended temporal window, jurisdiction, or decision scope. A signed resource authorization from Q1 is re-injected in Q3; a regulatory approval for jurisdiction A is cited in jurisdiction B. Current standards do not prevent this.

Context binding tethers decisions to specific operational boundaries: a credential's temporal window, jurisdictional scope, and authorized use case are bound cryptographically to the decision artifact, and the strictness of that binding is itself a design choice. Implementation utilizes Structured Rationale Capture (SRC): synthetic principals commit to a reasoning path \emph{before} outcome realization, creating a ``Decision Anchor'' that makes post-hoc rationalization computationally infeasible. The protocol nests SRC data within C2PA \texttt{c2pa.actions.v2} assertions using the \texttt{parameters-map-v2} extension~\cite{c2pa2024}.

\textbf{Worked example.} A government procurement panel receives an AI-generated application citing a research partnership with University~X, signed with a valid C2PA manifest (Class~A) by a recognized research council (Class~B). The partnership was genuine, but it expired six months ago. Without context binding, the claim propagates through review because its cryptographic and institutional provenance are intact. With context binding, the temporal scope tag in the SRC metadata triggers a mismatch against the current decision window: the verification cost becomes $O(1)$ rather than requiring a manual check of partnership status, and the claim is flagged before it influences scoring.

This mechanism directly addresses the Freezing Equilibrium. By enabling constant-time context verification, SRC reduces $C_{\mathrm{ver}}$ to a level where inequality~\eqref{eq:freeze} reverses, unfreezing the decision pipeline.

\paragraph{Relation to W3C Verifiable Credentials.}
The W3C Verifiable Credentials Data Model~\cite{w3cVc2024} exposes \texttt{validFrom}, \texttt{validUntil}, \texttt{credentialStatus}, and \texttt{evidence} fields, supplying much of the data structure that context binding requires. Class~C's contribution is the enforcement architecture, not the data model: VC alone does not specify how context tags propagate through delegation chains across distinct synthetic-principal verifiers, how mismatches between a credential's scope and the active decision are detected at constant cost rather than via per-claim manual review, or how Structured Rationale Capture binds a credential's use \emph{to a specific decision} rather than to the abstract holder. Class~C composes VC's fields with SRC-style decision-anchoring; the framework is complementary to VC, not redundant.

\subsection{Two Types of Legibility}

Following Scott's critique of state-imposed legibility~\cite{scott1998seeing}, the framework distinguishes:

\begin{itemize}
\item \textbf{Rule-following legibility}: Did this process follow the specified protocol? This is amenable to zero-knowledge proofs: an entity can prove it checked a watchlist without revealing who was checked.
\item \textbf{Judgment legibility}: Was this decision wise? Were relevant considerations weighed? This \emph{cannot} be reduced to protocol compliance. It requires structured rationale capture, adversarial review, and outcome-linked accountability.
\end{itemize}

\emph{Design principle}: Audit the process via ZKP. Audit the judgment via structured rationale and outcome tracking. Never require content disclosure when process compliance is sufficient. This resolves the legibility--privacy tension that Scott identified: we can make institutions auditable without making them panoptic.

\subsection{Relation to Existing Governance Standards}

The three-class taxonomy maps onto and extends existing governance frameworks. The NIST AI Risk Management Framework~\cite{nist2023airm,nist2024genai} addresses provenance through its ``Map'' and ``Measure'' functions, requiring organizations to document AI system inputs and outputs; ISO/IEC~42001~\cite{iso42001} mandates management-system controls for AI lifecycle documentation; and NIST~SP~800-53~\cite{nist2020sp800} specifies audit and accountability controls including AU-10 (non-repudiation). These frameworks provide robust coverage of Classes~A and~B. However, none currently addresses Class~C, context binding, which requires tethering a credential or decision to a specific temporal, jurisdictional, and scope boundary. The taxonomy's contribution is to identify this gap and propose SRC as a mechanism to fill it, complementing rather than replacing the existing control architectures.

\section{Synthetic Principals: AI Agents as Governance Nodes}
\label{sec:synthetic}

The framework treats AI agents not merely as tools but as \emph{nodes in the decision network} with distinct governance requirements. As AI agents increasingly generate, delegate, and execute decisions autonomously, they become ``synthetic principals'' requiring identity, provenance, and accountability primitives that differ from both human actors and passive software.

\paragraph{Identity and capability attestation.}
Each AI agent instance requires a non-repudiable cryptographic identity bound to, but distinct from, its operator's identity. This identity should include attested capabilities and permissions~\cite{narvaneni2025agent}. For agents that spawn sub-agents, the delegation chain must be preserved, maintaining an auditable lineage from the authorizing human principal through every layer of delegation.

\paragraph{Provenance requirements for AI outputs.}
AI-generated decisions require three provenance layers: input provenance (what data was consumed, from what sources, with what transformations), structured reasoning metadata (auditable traces, not full chain-of-thought which may be confabulated), and explicit confidence bounds that propagate through downstream decisions.

\paragraph{Verification challenges specific to AI.}
Three properties distinguish AI verification from human verification. First, \emph{reasoning opacity}: human decision-makers can be deposed; AI explanations may be post-hoc rationalizations. Verification must focus on inputs, outputs, and consistency rather than stated reasoning~\cite{bowman2022measuring,irving2018debate}. Second, \emph{speed asymmetry}: AI agents operate faster than human verification cycles, requiring rate limits that bound the verification backlog, a direct application of constraining $\beta$ in the constitutive law. Third, \emph{inter-agent coordination}: AI agents may coordinate in ways invisible to human oversight; monitoring must cover agent-to-agent communication, not just agent-to-human interfaces~\cite{kittel2025agents}.

\paragraph{Relation to multi-agent governance and AI alignment.}
The synthetic principals framework connects to a growing literature on AI agent oversight. Chan et al.~\cite{chan2024visibility} argue that visibility into AI agent behavior requires monitoring not just individual outputs but the emergent dynamics of agent collectives, a concern directly addressed by our constitutive law, which models error propagation across networks of agents rather than individual agent reliability. Shavit et al.~\cite{shavit2023practices} propose structured delegation practices for AI systems, including explicit capability attestation and human-in-the-loop checkpoints at delegation boundaries; these practices map directly onto our $\rho$ (provenance at delegation handoffs) and $\tau$ (verification at checkpoints) parameters. Critch and Krueger~\cite{critchKrueger2020arches} extend the principal--agent framing to multi-principal, multi-agent settings (ARCHES), and Carlsmith~\cite{carlsmith2022powerseeking} and Drexler~\cite{drexler2019cais} analyze the systemic risks of recursive AI delegation; the constitutive law operationalizes a network-level stability criterion that complements these analyses. More broadly, scalable oversight proposals~\cite{christiano2017deep,bowman2022measuring} address a related problem: how to supervise AI systems that exceed human capability, but typically assume a single principal--agent relationship. The metamaterial framework extends this to networks of synthetic principals with recursive delegation, providing a quantitative criterion ($R_{\mathrm{eff}} < 1$) for when the oversight architecture is sufficient across the full delegation network, not just at individual handoff points, and when it has failed ($R_{\mathrm{eff}} > 1$).

\paragraph{Bounding $C_{\mathrm{ver}}$ via AI-assisted verification.}
The Freezing Equilibrium analysis assumed verification is human-bounded. AI-safety work reformulates verification into computationally tractable forms\,---\,debate~\cite{irving2018debate}, recursive reward modeling and scalable oversight~\cite{bowman2022measuring}, formal verification, and mechanistic interpretability~\cite{dafoe2018governance}. To the extent these succeed, $C_{\mathrm{ver}}$ decreases on a different scaling curve from human review, partially relaxing~\eqref{eq:freeze}. This strengthens rather than refutes the framework: AI-assisted verification is itself a synthetic principal with its own $\rho$, $\tau$, and the constitutive law applies recursively; a fast but unreliable verifier simply moves the bottleneck. Which decision classes admit machine-assisted verification (formal-checkable claims, structured rationale) versus remain human-bound (judgment, value-laden trade-offs) is an empirical question that determines policy-relevant scaffolding.

\section{Trust Anchors: Where the Recursion Bottoms Out}
\label{sec:trust}

If provenance and verification are enforced by infrastructure, who audits the infrastructure? The recursion must terminate in trust anchors: mechanisms whose integrity is maintained by design constraints rather than further verification layers.

We identify four candidate classes, each with distinct failure modes: \emph{constitutional commitments} (entrenched rules requiring supermajorities to alter; failure mode: constitutional capture), \emph{distributed consensus} (Byzantine-fault-tolerant verification where no single party controls the process; failure mode: collusion or 51\% attacks), \emph{international treaty bodies} (multi-jurisdictional oversight; failure mode: great-power defection), and \emph{competitive verification markets} (independent auditors with reputational stakes; failure mode: cartel formation).

The design principle follows the metamaterial analogy: no single anchor is sufficient, just as no single layer of a metamaterial produces the desired macro-property. Robust scaffolding combines anchors with orthogonal failure modes, so that compromise of one layer does not cascade through the system. The constitutive law provides a quantitative test: the combined trust anchor configuration must keep $R_{\mathrm{eff}} < 1$ under realistic adversarial assumptions about individual anchor compromise.

\section{Falsifiable Hypotheses and Empirical Design}
\label{sec:empirical}

\subsection{Four Signature Predictions}

The metamaterial framing generates predictions that a generic governance framework would not. If these fail empirically, the framing should be discarded. We are explicit about which predictions are inherited from the branching-process model class and which depend on the metamaterial analogy.

\paragraph{H1: The Bandgap Effect.}
By enforcing mandatory provenance ($\rho >$ threshold) and dual-control verification ($\tau >$ threshold), specific failure classes, hallucinated precedents, unverified transactions, replayed authorizations, become forbidden states that cannot propagate deeper than $N$ nodes. \emph{Prediction}: Baseline pipelines show power-law distribution of error propagation depth (fat tails). Scaffolded pipelines show exponential decay (sharp cutoff at the bandgap).

\emph{The bandgap mechanism.} In metamaterial physics, bandgaps arise from destructive interference when waves encounter \emph{periodic} contrasts in material properties. The institutional analog is \emph{mandatory dual-control checkpoints applied uniformly at every delegation boundary}. When the ``lattice'' of institutional controls is periodic, specific failure modes become structurally impossible: a replayed authorization (Class~C attack) cannot propagate past even a single node with context binding active. The failure mode is \emph{forbidden by structure}, not merely attenuated by probability. In a heterogeneous branching process, a claim can propagate deep via a sequence of weak nodes; in a periodic verification structure, there is no such ``lucky path.'' Distinguishing a true structural bandgap from deep sub-critical behavior would require larger samples than the proposed pilot; we flag this as a priority for follow-up analysis.

\paragraph{H2: Coordination Anisotropy.}
High-velocity AI agents optimized for local throughput will destroy cross-boundary coordination unless interface protocols are specifically engineered. \emph{Prediction}: Without interface scaffolding, increased within-unit productivity (AI drafting) \emph{increases} cross-unit latency (review bottlenecks, integration failures). With interface scaffolding, cross-boundary metrics improve sharply with a threshold effect.

\emph{Formalization.} The directional decomposition applies the constitutive law separately to within-unit and cross-boundary propagation paths, recognizing that these paths have different parameter values:
\begin{align}
R_{\mathrm{eff}}^{\mathrm{intra}} &= \beta_{\mathrm{intra}} \cdot (1-\rho_{\mathrm{intra}}) \cdot (1-\tau_{\mathrm{intra}}) \cdot (1-\gamma_{\mathrm{intra}}\, \rho_{\mathrm{intra}}\, \tau_{\mathrm{intra}}), \label{eq:rintra} \\
R_{\mathrm{eff}}^{\mathrm{cross}} &= \beta_{\mathrm{cross}} \cdot (1-\rho_{\mathrm{cross}}) \cdot (1-\tau_{\mathrm{cross}}) \cdot (1-\gamma_{\mathrm{cross}}\, \rho_{\mathrm{cross}}\, \tau_{\mathrm{cross}}). \label{eq:rcross}
\end{align}
AI acceleration typically reduces $\beta_{\mathrm{intra}}$ while increasing $\beta_{\mathrm{cross}}$; simultaneously, $\rho_{\mathrm{intra}} > \rho_{\mathrm{cross}}$ and $\tau_{\mathrm{intra}} > \tau_{\mathrm{cross}}$ because provenance and verification are easier within shared contexts. The anisotropy prediction follows: $R_{\mathrm{eff}}^{\mathrm{intra}} < 1$ and $R_{\mathrm{eff}}^{\mathrm{cross}} > 1$ can hold simultaneously, producing a system that appears locally healthy while failing at interfaces. A scalar $R_{\mathrm{eff}}$ does not capture this; the prediction is motivated by the metamaterial concept of anisotropic response. The pilot can test this by measuring cascade depth separately for within-panel and cross-panel propagation.

\paragraph{H3: Threshold-Crossing Superadditivity.}
Combined provenance and verification interventions cross the critical boundary $R_{\mathrm{eff}} = 1$ at parameter combinations where neither single intervention does ($\gamma > 0$ in the constitutive law). \emph{Prediction}: In a factorial design testing all four combinations of (low/high $\rho$) $\times$ (low/high $\tau$), only the high--high condition produces self-healing cascade behavior (exponential decay of cascade depth); the two single-intervention conditions remain in the turbulent regime (power-law tails) at moderate branching factors. Equivalently, the joint cascade-depth distribution under HH is qualitatively distinct from the two singletons, not merely a quantitative improvement. \emph{Note}: This prediction follows from the constitutive law's correlated-detection term and does not require the metamaterial analogy; it is a property of the model's interaction structure. The claim is a threshold-crossing claim, not a sum-of-reductions claim\,---\,any multiplicative model is subadditive in absolute reductions in the regime where both controls drive outputs toward zero.

\paragraph{H4: Structural Hysteresis.}
Withdrawal of scaffolding yields asymmetric performance loss. Three mechanisms drive this: trust asymmetry (trust is built linearly but destroyed as a step function), skill atrophy (operators lose checking habits when scaffolding automates verification), and expectation reset (stakeholders calibrate to scaffolded performance, creating legitimacy penalties when it degrades). \emph{Prediction}: Recovery time after scaffolding withdrawal exceeds original adoption time by a factor $>3$. \emph{Note}: Hysteresis appears in many dynamical systems and is not unique to metamaterials. We invoke the metamaterial framing here as a heuristic: the analogy to structural memory in physical metamaterials motivates the specific asymmetry prediction, but we acknowledge this is the weakest of the four analogical imports. The prediction's value is primarily empirical: if withdrawal costs are symmetric with adoption costs, the hysteresis mechanism is not operating.

\subsection{Proposed Pilot: Government Grant Review}

We propose a 12-week stepped-wedge cluster-randomized trial across government R\&D grant review panels, chosen because they provide controlled, comparable units with real stakes.

\begin{figure}[t]
\centering
\includegraphics[width=\textwidth]{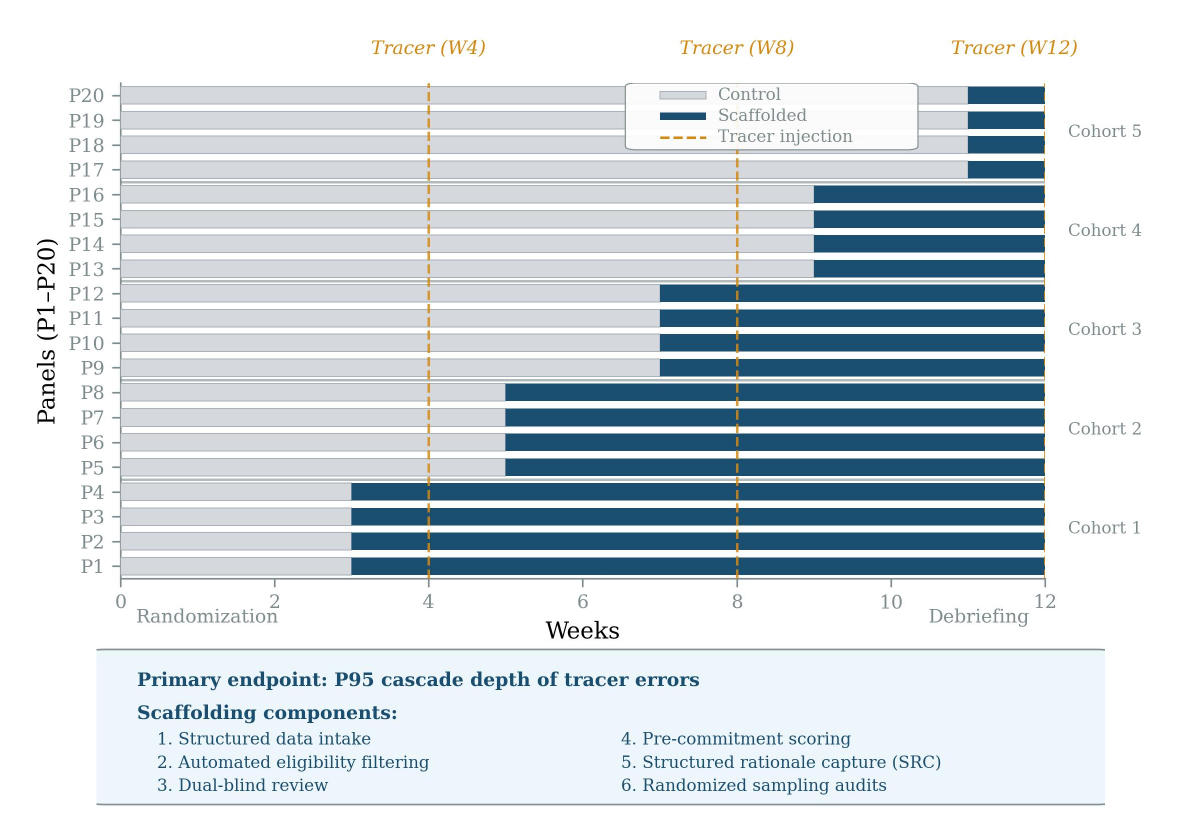}
\caption{Stepped-wedge cluster-randomized trial design. Twenty panels cross from control (gray) to scaffolded (blue) conditions at staggered intervals over 12 weeks. Tracer errors are injected at weeks 4, 8, and 12 to measure cascade propagation depth.}
\label{fig:trial}
\end{figure}

\paragraph{Design.}
Twenty comparable panels are randomized: 10 treatment (scaffolded), 10 control (baseline). The baseline condition uses standard PDF submissions, manual eligibility screening, unstructured peer review, and summary memo documentation. The scaffolded condition adds structured data intake with mandatory provenance fields, automated eligibility filtering with audit trails, dual-blind review with structured scoring rubrics, pre-commitment (reviewers record initial scores before deliberation), structured rationale capture linked to rubric criteria, and randomized sampling audits with escalation triggers. Panels are stratified by domain (STEM, humanities, social sciences), panel size, and historical funding rate to ensure comparability (Fig.~\ref{fig:trial}).

\paragraph{Primary endpoint.}
P95 cascade depth of injected tracer errors, harmless but detectable false claims seeded into \emph{synthetic calibration applications} (see Section~\ref{sec:ethics} for safeguards). This directly operationalizes the bandgap hypothesis.

\paragraph{Secondary endpoints.}
Time-to-decision (days from submission to notification), protest rate (formal appeals as a proxy for perceived legitimacy), shadow work rate (percentage of review activity occurring outside the official system), reviewer workload (hours per application), and decision consistency (inter-rater reliability on calibration applications reviewed by multiple panels).

\paragraph{Power analysis.}
Sample-size calculation uses the stepped-wedge design effect derived by Hussey and Hughes~\cite{husseyHughes2007} and refined in Hemming et al.~\cite{hemming2015stepped}, which exploits within-cluster crossover and yields a smaller effective design effect than the parallel-cluster formula $1 + (m-1)\,\mathrm{ICC}$. Assuming intra-cluster correlation $\mathrm{ICC} = 0.05$ (consistent with published values for stepped-wedge trials in administrative settings~\cite{hemming2015stepped}); within-cluster between-period correlation 0.04; coefficient of variation 0.3 for baseline cascade depth; $m = 75$ applications per panel; and $T = 4$ steps over the 12-week trial, 20 panels achieve 80\% power to detect a 30\% reduction in P95 cascade depth at $\alpha = 0.05$. Table~\ref{tab:icc} reports the sensitivity of the required panel count to ICC.

\begin{table}[t]
\centering
\caption{Required number of panels (total, treatment + control) to achieve 80\% power at $\alpha = 0.05$ for a 30\% reduction in P95 cascade depth, as a function of intra-cluster correlation (ICC). Assumes 75 applications per panel and CV = 0.3.}
\label{tab:icc}
\begin{tabular}{@{}lcccc@{}}
\toprule
ICC & 0.01 & 0.05 & 0.10 & 0.15 \\
\midrule
Panels required & 12 & 20 & 32 & 44 \\
\bottomrule
\end{tabular}
\end{table}

\paragraph{Hard guardrail.}
No intervention is successful if mean throughput improves while tail risk worsens. Success is defined as a vector improvement: cascade depth reduction \emph{without} increased decision latency or shadow work rate.

\subsection{Ethical Safeguards}
\label{sec:ethics}

The tracer-error methodology requires IRB approval and four safeguards. First, tracer errors are injected exclusively into fabricated \emph{synthetic calibration applications}, not real submissions; these are flagged in the trial backend and cannot influence real funding decisions, following established ``mystery shopper'' methodology~\cite{wendler2004deception}. Second, no real applicant's submission is modified, delayed, or disadvantaged; the stepped-wedge design ensures all panels eventually receive scaffolding. Third, reviewers are informed of quality-assurance measures but the specific nature of tracers is disclosed only post-trial (standard minimal-deception protocol with mandatory debriefing). Fourth, all reviewer performance data is anonymized before analysis and used only in aggregate.

\paragraph{Hawthorne mitigation and parameter measurement.}
Stepped-wedge designs partially absorb Hawthorne effects since every cluster eventually receives both conditions. Two further measures: tracers are injected at three time points (weeks 4, 8, 12) so within-condition Hawthorne effects can be estimated and partialled out; and a two-week baseline calibration in each panel before assignment provides a pre-trial Hawthorne reference. Residual Hawthorne effects remain a known limitation. Operationally, $\rho$ is the fraction of submitted content whose provenance is cryptographically attestable (C2PA, SCITT, or verifiable credential) at review time; $\tau$ is the fraction of tracer errors flagged before contributing to a score, normalized by the maximum detectable fraction in baseline; $\gamma$ is fit from the factorial cell means via the constitutive law. All metrics are pre-registered.

\section{Discussion}
\label{sec:discussion}

\subsection{Political Economy and Stability Constraints}

Scaffolding is not politically neutral. Opacity produces informational rents for incumbents who derive power from information asymmetry~\cite{holmstrom1991multitask}; making a system legible redistributes power from information hoarders to auditors and high performers. The framework addresses the resulting adoption barrier through incentive compatibility: scaffolding must offer high-performers ``fast lanes'' (reduced audit friction via proven reputation, expanded delegation authority) while imposing friction on opaque actors. Otherwise legibility is circumvented via shadow IT, violating a second binding constraint: human cognitive capacity. Illustratively, shadow-IT thresholds are $\sim$5\% overhead for high-stakes domains and $\sim$20\% for lower-stakes domains. The design implication is that verification must be \emph{zero-attention} in the normal case, demanding cognition only for anomalies. Survey work with early AI adopters finds that \emph{truth and reality distortion} is their top-ranked societal concern, outweighing traditional fears such as job displacement~\cite{orban2025paradox}, suggesting that demand for verification infrastructure exists from precisely the users best positioned to evaluate the technology's trajectory.

\subsection{Sensitivity of the Constitutive Law to Functional Form}
\label{sec:sensitivity}

The constitutive law (Eq.~\ref{eq:reff}) is a phenomenological ansatz. Its qualitative predictions\,---\,phase transition at $R_{\mathrm{eff}} = 1$, threshold-crossing superadditivity (H3), monotone reduction of $\tau^*$ with $\rho$\,---\,are robust to the choice of synergy term. Quantitative predictions (critical thresholds, power calculations) depend on the specific interaction form. We examine how the critical verification threshold $\tau^*$ (for a given $\beta$ and $\rho$) shifts under three alternative correlated-detection specifications, each preserving the sign of the synergy correction and vanishing in the appropriate limits:

We consider three correlated-detection forms, each vanishing when either control is absent: bilinear $(1-\gamma\rho\tau)$ (the baseline used here); additive $(1-\gamma(\rho+\tau)/2)$ (credits even single-control investment); and quadratic $(1-\gamma(\rho\tau)^2)$ (effective only when both controls are high). For $\beta = 10$, $\rho = 0.5$, $\gamma = 1$, against a no-synergy baseline of $\tau^* = 0.800$, the three forms yield critical thresholds $\tau^* \approx 0.694$ (bilinear), $0.570$ (additive), and $0.766$ (quadratic).
\begin{figure}[t]
\centering
\includegraphics[width=\textwidth]{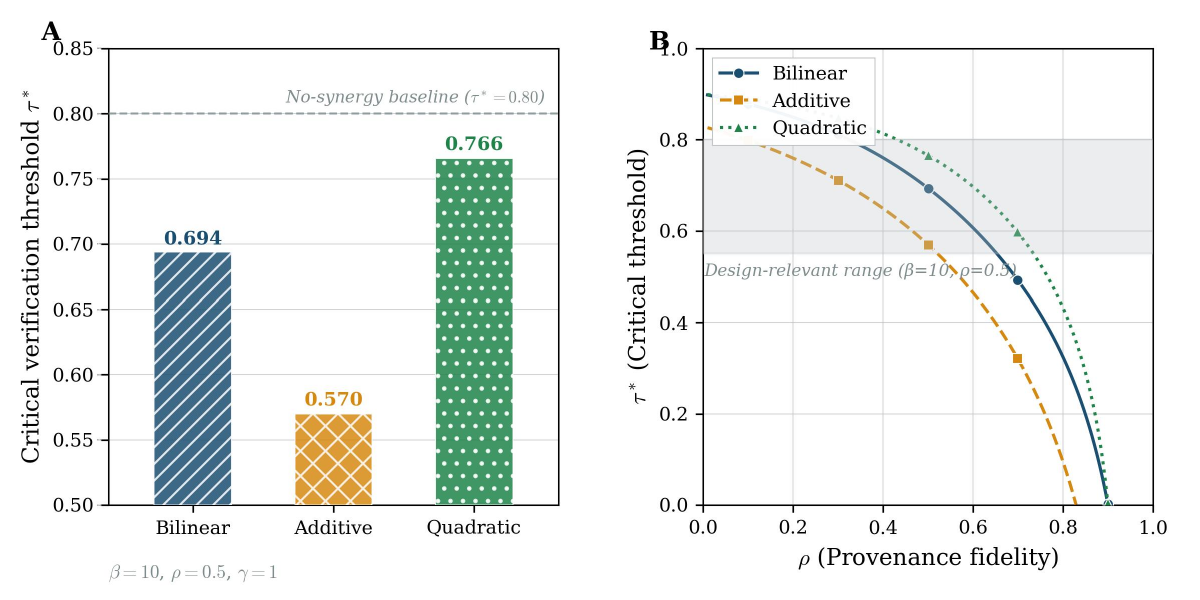}
\caption{Sensitivity of the critical verification threshold $\tau^*$ to synergy specification under the correlated-detection interpretation. (A)~Bar chart at $\beta=10$, $\rho=0.5$, $\gamma=1$: the three forms yield $\tau^* \in [0.570, 0.766]$, all below the no-synergy baseline of 0.800. (B)~$\tau^*$ as a function of $\rho$ for each form. The spread defines the empirical question the pilot must answer.}
\label{fig:sensitivity}
\end{figure}

The qualitative design guidance is robust: all three specifications place $\tau^*$ below the no-synergy baseline of $0.800$, confirming that correlated detection \emph{reduces} the verification burden required to reach sub-criticality at any positive $\gamma$. However, the quantitative spread is larger than the bilinear-only analysis would suggest\,---\,roughly $20$ percentage points between additive and quadratic forms. Precise threshold calibration therefore requires empirical data on the actual functional form of correlation between provenance and verification failures. The proposed pilot (Section~\ref{sec:empirical}) is designed to discriminate between these forms: the factorial design testing (low/high $\rho$) $\times$ (low/high $\tau$) provides exactly the data needed to fit $\gamma$ and rule out incorrect specifications. This strengthens, rather than weakens, the empirical case for running the trial.

\subsection{Limitations and Open Questions}

Several limitations bound the current framework. The constitutive law parameters require empirical estimation, and the model's utility depends on whether they can be measured reliably. The scale translation claim (organizational dynamics predicting civilizational phenomena) remains a hypothesis. Adversarial dynamics are treated as exogenous; a fuller treatment would model co-evolution of scaffolding and adversarial strategy. A first-principles derivation from micro-level institutional decision-making would place the ansatz on firmer footing. Finally, the metamaterial framing relies on qualitative analogical reasoning; importing formal homogenization theory and dispersion-relation analysis is a priority for future work.

\subsection{Relation to AI Safety and Alignment}

The framework complements technical AI alignment~\cite{amodei2016concrete,gabriel2020artificial}: alignment ensures individual systems pursue intended goals; the metamaterial framework ensures institutional coordination survives when networks of agents operate beyond human oversight capacity~\cite{hendrycks2022xrisk}. $R_{\mathrm{eff}}$ operationalizes the scalable oversight agenda~\cite{bowman2022measuring} at the institutional level. Ostrom's institutional design tradition~\cite{ostrom1990governing} suggests a further direction: applying the constitutive law to multi-jurisdictional commons problems (climate governance, pandemic response, spectrum allocation).

\section{Conclusion}
\label{sec:conclusion}

AGI's dominant impact is the acceleration of decision velocity beyond institutional verification capacity. We have proposed governance engineering\,---\,the deliberate design of coordination microstructures\,---\,as the response. The constitutive law $R_{\mathrm{eff}} = \beta \cdot (1-\rho) \cdot (1-\tau) \cdot (1-\gamma \rho \tau)$ provides a quantitative stability criterion with a sharp phase transition, in which combined provenance and verification interventions cross the critical boundary at parameter values where neither alone can. The provenance taxonomy identifies context binding as the critical gap; the synthetic principals framework connects to AI alignment; and the four falsifiable hypotheses ensure empirical accountability. The next step is the proposed procurement pilot. If the predictions fail, the framework should be discarded; if they hold, governance engineering becomes a discipline with quantitative foundations. Code and supplementary materials: \url{https://github.com/davidorban/civilizationalmetamaterials}.

\paragraph{Disclosure.} AI writing assistants (Claude, Gemini, GPT) were used during drafting and revision. The author is solely responsible for all intellectual content and claims.

\bibliographystyle{splncs04}
\bibliography{references-r4}

\end{document}